\documentclass[conference]{IEEEtran}
\usepackage{graphicx}
\usepackage{amsmath,bbm,amssymb,amsfonts,amstext,amsopn,sansmath}
\usepackage{cite}
\usepackage{balance}
\usepackage{url}
\usepackage{epsfig}
\usepackage{setspace}
\usepackage{stmaryrd}
\usepackage{psfrag}	
\usepackage{multirow}
\usepackage{float}
\usepackage[process=auto]{pstool}
\usepackage{etoolbox}
\usepackage{algorithm}
\usepackage{algorithmic}
\allowdisplaybreaks
\usepackage[nolist]{acronym}

\newcommand{\Trans}{^{\mathsf{T}}}
\newcommand{\Herm}{^{\mathsf{H}}}

\newcommand{\Ptx}{P_{\mathrm{tx}}}
\newcommand{\Prd}{P_{\mathrm{rd}}}
\newcommand{\Ptot}{P_{\mathrm{tot}}}

\newcommand{\Pbb}{P_{\mathrm{bb}}}

\newcommand{\Pamp}{P_{\mathrm{amp}}}
\newcommand{\Prfc}{P_{\mathrm{rfc}}}

\newcommand{\Hc}{\mathbf{H}}

\newcommand{\Ex}{\mathbbmss{E}}

\newtheorem{prop}{Proposition}

\newtoggle{OneColumn}

\toggletrue{OneColumn}

\iftoggle{OneColumn}{%
  
}{%
}

\EndPreamble
\begin{document}
\title{Scalable and Energy-Efficient Millimeter Massive MIMO Architectures: \\ Reflect-Array and Transmit-Array Antennas}

\author{
Vahid Jamali, Antonia M. Tulino,  Georg Fischer, Ralf M\"uller, and Robert Schober
}

\maketitle

\begin{abstract}
Hybrid analog-digital architectures are considered as promising candidates for implementing millimeter wave (mmWave) massive multiple-input multiple-output (MIMO) systems since they enable a considerable reduction of the required number of costly radio frequency (RF) chains by moving some of the
signal processing operations into the analog domain. However, the analog feed network, comprising RF dividers, combiners, phase shifters, and line connections, of hybrid MIMO architectures is not scalable due to its prohibitively high power consumption for large numbers of transmit antennas. Motivated by this limitation, in this paper, we study novel massive MIMO architectures, namely reflect-array (RA) and transmit-array (TA) antennas. We show that the precoders for RA and TA antennas have to meet different constraints compared to those for conventional MIMO architectures. Taking these constraints into account and exploiting the sparsity of mmWave channels, we design an efficient precoder for RA and TA antennas based on the orthogonal matching pursuit algorithm. Furthermore, in order to fairly compare the performance of RA and TA antennas with conventional fully-digital and hybrid MIMO architectures, we develop a unified power consumption model.
Our simulation results show that unlike conventional MIMO architectures, RA and TA antennas are highly energy efficient and fully scalable in terms of the number of transmit antennas.
\end{abstract}

\acresetall
\section{Introduction}

Millimeter wave (mmWave) communication systems are promising candidates to meet  the high data rate requirements of the next generation of wireless communication networks \cite{gao2018low,ahmed2018survey}.
These systems are typically assumed to be equipped with a large array of antennas at the transmitter and/or the receiver to cope with the high path loss, limited scattering, and small antenna apertures at mmWave frequencies. However, conventional fully-digital (FD) multiple-input multiple-output (MIMO) systems, in which each antenna is connected to a dedicated radio frequency (RF) chain, are infeasible for mmWave systems due to the prohibitively high cost and energy consumption of high resolution analog-to-digital/digital-to-analog converters \cite{gao2018low}. This has motivated researchers to consider hybrid analog-digital MIMO architectures, which tremendously reduce the required number of RF chains by moving some of the signal processing operations into the analog domain \cite{ahmed2018survey,el2014spatially}.

Typically, in hybrid MIMO systems, it is assumed that the output of each RF chain is connected to all antennas. This architecture is referred to as fully-connected (FC) hybrid MIMO and is able to realize the full beamforming gain of massive antenna arrays. Unfortunately, FC hybrid MIMO is not scalable due to the excessive power consumption of the analog feed network for large numbers of antennas \cite{yan2018performance}. In particular, the analog feed network is comprised of RF dividers, combiners, phase shifters, and line connections, which consume huge amounts of power and hence reduce energy efficiency. To deal with this issue, partially-connected (PC) hybrid MIMO architectures were considered in the literature where the output of each RF chain is connected to only a subset of the antennas \cite{gao2018low,gao2016energy}. Thereby, no RF combiner is needed, and the numbers of required phase-shifters and RF lines are reduced. Nevertheless, as will be shown in this paper, the power consumption of PC hybrid MIMO still scales with the number of antennas in a similar manner as FC hybrid~MIMO.

\begin{figure}
  \centering
 \scalebox{0.5}{
\pstool[width=2\linewidth]{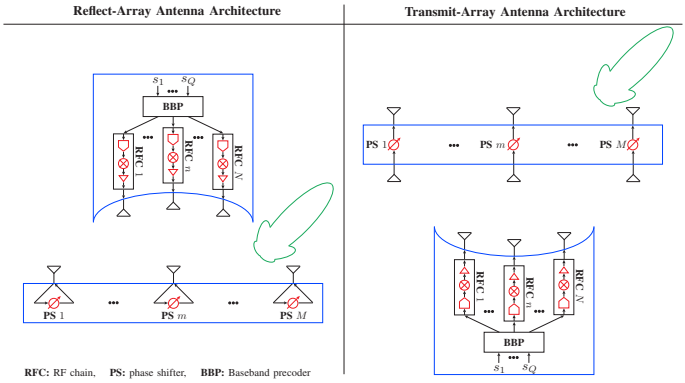}{
\psfrag{A1}[c][c][0.8]{\textbf{PS}~$1$}
\psfrag{A3}[c][c][0.8]{\textbf{PS}~$m$}
\psfrag{A5}[c][c][0.8]{\textbf{PS}~$M$}
\psfrag{B1}[c][c][0.8]{\textbf{RFC $1$}} 
\psfrag{B2}[c][c][0.8]{\textbf{RFC $n$}}
\psfrag{B3}[c][c][0.8]{\textbf{RFC $N$}} 
\psfrag{C}[c][c][0.8]{\textbf{BBP}} 
\psfrag{D1}[c][c][1]{$s_1$}
\psfrag{D2}[c][c][1]{$s_Q$}
\psfrag{K4}[c][c][1]{\textbf{Reflect-Array Antenna Architecture}}
\psfrag{K5}[c][c][1]{\textbf{Transmit-Array Antenna Architecture}}
\psfrag{T}[l][c][0.8]{\textbf{RFC:} RF chain,$\quad$ \textbf{PS:} phase shifter,$\quad$ \textbf{BBP:} Baseband precoder}
}} 
\caption{Schematic illustration of the considered reflect-array and transmit-array massive MIMO architectures. }
\label{Fig:SysMod}
\end{figure}

In order to improve the scalability and energy-efficiency of mmWave massive MIMO systems, we consider two novel massive MIMO architectures in this paper, namely  reflect-array (RA) and transmit-array (TA) antennas, see Fig.~\ref{Fig:SysMod}. Both architectures comprise a large array of passive antenna elements and a few active antennas (usually horn antennas). Each active antenna is equipped with a dedicated RF chain and illuminates the array of passive antennas. Therefore, each passive element receives a superposition of the signals transmitted (over the air) by the active antennas and adds a desired phase shift to the overall signal. In RA, the phase-delayed signal is then reflected from the array whereas in TA, the phase-delayed signal is transmitted in the forward direction\footnote{RA and TA antennas have several advantages/disadvantages with respect to each other. For instance, in RA, the feed position introduces a blocking area whereas this issue does not exist in TA antennas. On the other hand, RA systems facilitate the placement of the control system for the phase shifters on the back side of the array \cite{abdelrahman2017analysis}.}. Borrowing an analogy from optics, an RA is analogous to a curved mirror  whereas a TA is analogous to a lens. RA and TA antennas have been widely investigated in the microwave and antennas community and prototypes are available in the literature \cite{berry1963reflectarray,pozar1997design,abdelrahman2017analysis,lau2011transmitarray,di2015reconfigurable}. Thereby, the performance of these architectures is typically characterized in terms of the beamforming gain. In contrast, in this paper, we are interested in multiplexing several data streams and the design of the corresponding precoder. 

In this paper, we study RA and TA antennas and show that their corresponding precoders have to meet different constraints compared to those for conventional MIMO architectures. Taking these constraints into account and exploiting the sparsity of mmWave channels, we design a precoder for RA and TA antennas based on orthogonal matching pursuit (OMP). Furthermore, in order to fairly compare the performance of RA and TA antennas with conventional fully-digital and hybrid MIMO architectures, we develop a unified power consumption model which includes the impacts of the loss over air for the RA and TA architectures, the RF feed network for the FC and PC hybrid architectures, and the digital processing and power amplifiers for all architectures. Our simulation results show that in contrast to the conventional FD, FC, and PC MIMO architectures,  the RA and TA MIMO architectures are highly energy-efficient and fully scalable in terms of the number of transmit antennas. We note that the recent paper \cite{zhou2018hardware} also studied RA antennas where a precoder was designed based on alternating optimization (AO). We employ this precoder as a benchmark and show that the proposed OMP-based precoder outperforms the AO-based precoder in \cite{zhou2018hardware}. Moreover, the focus of this paper is mainly the scalability and energy-efficiency of RA and TA MIMO systems which was not studied in \cite{zhou2018hardware}. Furthermore, in this paper, a more detailed model for the channel between the active and passive antennas (which affects the model for the precoder structure) is considered compared to \cite{zhou2018hardware}.

\textit{Notations:} Bold capital and small letters are used to denote matrices and vectors, respectively. $\|\mathbf{A}\|_F$, $\mathbf{A}\Trans$, and $\mathbf{A}\Herm$ denote the Frobenius norm, transpose,  and Hermitian  of matrix $\mathbf{A}$, respectively. $\Ex\{\cdot\}$ represents expectation and $\angle a$ is the angle of the complex number $a$ in polar coordinates. Moreover,  $\mathcal{CN}(\boldsymbol{\mu},\boldsymbol{\Sigma})$ denotes a complex normal random variable (RV) with mean vector $\boldsymbol{\mu}$ and covariance matrix $\boldsymbol{\Sigma}$. Furthermore, $\boldsymbol{0}_n$ and $\boldsymbol{0}_{n\times m}$ denote a vector of size $n$ and a matrix of size $n\times m$, respectively, whose elements are all zeros. Moreover,  $\mathbf{I}_n$ is the $n\times n$ identity matrix and $\mathbb{C}$ represents the set of complex numbers. $\mathrm{vec}(\mathbf{A})$ denotes the vectorized version of matrix $\mathbf{A}$. Moreover, $[a(m,n)]_{m,n}$ represents a matrix with element $a(m,n)$ in its $m$-th row and $n$-th column. $\mathbf{A}_{m,n}$ and $\mathbf{a}_n$ denote the element in the $m$-th row and $n$-th column of matrix $\mathbf{A}$ and the $n$-th element of vector $\mathbf{a}$, respectively. Finally, $\mathrm{vec}(\mathbf{A})$ returns a vector whose elements are the stacked columns of matrix~$\mathbf{A}$.

\section{System, Channel, Signal, and Power Consumption Models}
In this section, we present the system and channel models as well as the signal and power consumption models.

\subsection{System and Channel Models}
We consider a point-to-point MIMO system where the transmitter and receiver have $M$ and $J$ antennas, respectively. The input-output MIMO model is given by
\begin{IEEEeqnarray}{lll} \label{Eq:AWGN}
\mathbf{y} = \Hc\mathbf{x}+\mathbf{z},
\end{IEEEeqnarray}
where $\mathbf{x}\in\mathbb{C}^{M\times 1}$ and $\mathbf{y}\in\mathbb{C}^{J\times 1}$ are the transmit and received vectors, respectively. Moreover, $\mathbf{z}\in\mathbb{C}^{J\times 1}$ denotes the additive white Gaussian noise vector at the receiver, i.e., $\mathbf{z}\sim\mathcal{CN}(\boldsymbol{0}_J,\sigma^2\mathbf{I}_J)$ where $\sigma^2$ denotes the noise variance at each antenna. Furthermore, $\Hc\in\mathbb{C}^{J\times M}$ is the channel matrix, which assuming the Saleh-Valenzuela model, is given by \cite{gao2016energy}
\begin{IEEEeqnarray}{lll} \label{Eq:Channel}
\Hc = \frac{1}{\sqrt{L}}\sum_{l=1}^L h_l \mathbf{h}_r(\theta_l^r,\phi_l^r)\mathbf{h}_t\Herm(\theta_l^t,\phi_l^t),
\end{IEEEeqnarray}
where $L$ is the number of effective channel paths corresponding to a limited number of scatterers and $h_l\in\mathbb{C}$ is the channel coefficient of the $l$-th path. Moreover, $\mathbf{h}_t(\theta_l^t,\phi_l^t)$ ($\mathbf{h}_r(\theta_l^r,\phi_l^r)$) denotes the transmitter (receiver) antenna array response vector at elevation angle $\theta_l^t\in[0,\pi]$ ($\theta_l^r\in[0,\pi]$) and azimuth angle $\phi_l^t\in[0,2\pi]$ ($\phi_l^r\in[0,2\pi]$). For a uniform planar transmit array, we can obtain $\mathbf{h}_t(\phi_l^t,\theta_l^t)$ as \cite{gao2016energy}
\begin{IEEEeqnarray}{lll} \label{Eq:ULA}
\mathbf{h}_t(\theta_l^t,\phi_l^t)  = \nonumber \\
\mathrm{vec}\left(\left[e^{j\frac{2\pi d}{\lambda}\left((m_1-1)\sin(\theta_l^t)\sin(\phi_l^t)+(m_2-1)\cos(\theta_l^t)\right)}\right]_{m_1,m_2}\right), \quad
\end{IEEEeqnarray}
where $\lambda$ denotes the wavelength, and $d$ is the distance between the array antenna elements. Assuming a uniform planar receiver array, $\mathbf{h}_r(\theta_l^r,\phi_l^r)$ can be modeled in a similar manner as $\mathbf{h}_t(\theta_l^t,\phi_l^t)$ in (\ref{Eq:ULA}).

\subsection{Transmit Signal and Power Consumption Models}\label{Sec:Power}

Let $\mathbf{s}\in\mathbb{C}^{Q\times 1}$ denote the vector of $Q$ independent data streams that we wish to transmit. Assuming linear precoding, the relation between $\mathbf{x}$ and $\mathbf{s}$ is as follows
\begin{IEEEeqnarray}{lll} \label{Eq:Sig_Gen}
\mathbf{x} =\sqrt{\Ptx}\mathbf{F}\mathbf{s},
\end{IEEEeqnarray}
where $\mathbf{F}\in\mathbb{C}^{M\times Q}$ is the precoder and $\Ptx$ denotes the transmit power. Here, we assume $\Ex\{\mathbf{s}\mathbf{s}\Herm\}=\mathbf{I}_Q$ and $\|\mathbf{F}\|_F = 1$~\footnote{In this paper, we consider a constraint on the maximum power radiated from the passive array which is typically enforced by regulations. Alternatively, one can consider a constraint on the power radiated from the active antennas. Although our derivations in Section~\ref{Sec:Model} and the proposed precoder in Section~\ref{Sec:Precoder} can be applied under both power constraints, we focus on the former power constraint for RA and TA antennas since this enables a more straightforward comparison with conventional MIMO architectures.}.  In order to fairly compare the power consumptions of the conventional MIMO architectures, i.e., FD, FC, and PC, and the proposed new MIMO architectures, i.e., RA and TA, a power consumption model that accounts for digital baseband processing, the RF network, and the power amplifiers is needed. 

\textbf{Baseband Circuitry:} The circuit power consumption comprises the power consumed for baseband processing, denoted by $\Pbb$, and by each RF chain (i.e., by the digital-to-analog converter, local oscillator, and mixer), denoted by $\Prfc$. Note that although in principle $\Pbb$ may vary as a function $M$, in the remainder of this paper, we assume  $\Pbb$ is constant since its impact is typically much smaller than that of $\Prfc$ \cite{lin2016energy}.

\textbf{RF Network:} In this paper, we assume an RF network with passive phase shifters, dividers, and combiners which introduce insertion loss. For large RF networks, the insertion loss may easily exceed $30$~dB which makes its precompensation infeasible due to amplifier nonlinearity at high gains \cite{yan2018performance}. In practice, to compensate for this insertion loss, several stages of power amplification are implemented throughout the RF network to ensure that a minimum power is delivered to drive the power amplifiers before transmission via the antennas.  For instance, for the gain-compensation amplifier design in \cite{yan2018performance}, each amplifier has up to $15$~dB gain at $40$~mW power consumption.  The number of required intermediate power amplifiers, denoted by $N_{\mathrm{amp}}$, depends on the specific RF network  architecture. Motivated by the experimental design in \cite{yan2018performance} and for analytical tractability, we assume three stages of power amplification where the signal is pre-amplified once before being fed to the power divider, passive phase shifter, and power combiner, respectively, to compensate the losses incurred in each stage. 

\textbf{Power Amplifiers:} The power consumed by the power amplifiers is commonly modeled as $\Prd/\rho_{\mathrm{pa}}$ where $\Prd$ is the radiated output power and  $\rho_{\mathrm{pa}}$ denotes the power amplifier efficiency \cite{yan2018performance,ribeiro2018energy,Rodriguez2016_MIMO_Loss,lin2016energy}.

In summary, the total power consumption, denoted by $\Ptot$, is obtained as 
\begin{IEEEeqnarray}{lll}\label{Eq:Power}
\Ptot =\Pbb+N\Prfc+N_{\mathrm{amp}}\Pamp+\Prd/\rho_{\mathrm{pa}},
\end{IEEEeqnarray}
where $\Pamp$ is the power consumed by each RF power amplifier used in the RF network. Note that for conventional MIMO architectures, $\Prd$ is identical to $\Ptx$, whereas for RA and TA antennas, $\Prd$ is the power radiated by the active antennas which is not necessarily the same as the power $\Ptx$ radiated by the passive array, cf. Section~\ref{Sec:RA_TA} for details.

\section{Mathematical Characterization of Different MIMO Architectures}\label{Sec:Model}
In this section, we characterize the constraints that different MIMO systems impose on precoder matrix $\mathbf{F}$ and the corresponding total power consumption $\Ptot$ as a function of $M$ and $N$. 

\subsection{Conventional MIMO Architectures}
In the following, we study the conventional FD, FC, and PC MIMO architectures. 

\subsubsection{Fully-Digital MIMO Architectures} Here, we have $N=M$ RF chains which enable FD precoding, i.e., $\mathbf{F}=\mathbf{B}$ where $\mathbf{B}$ is referred to as the digital precoder. Moreover, since we do not have an analog RF network, we obtain $N_{\mathrm{amp}}=0$. Therefore, the total consumed power is given by 
\begin{IEEEeqnarray}{lll} 
	\Ptot =\Pbb+M\Prfc+\Ptx/\rho_{\mathrm{pa}}.
\end{IEEEeqnarray}

\subsubsection{Fully-Connected Hybrid MIMO Architectures} In the FC hybrid architecture, we have $N$ RF chains whose outputs are connected to all $M$ antennas via analog dividers, phase shifters, and combiners. Typically, the relation $Q\leq N\ll M$ holds. For this MIMO architecture, the precoder has structure
\begin{IEEEeqnarray}{lll} \label{Eq:Sig_Full}
	\mathbf{F} = \mathbf{R}\mathbf{B},
\end{IEEEeqnarray}
where $\mathbf{B}\in\mathbb{C}^{N\times Q}$ denotes the digital precoder and $\mathbf{R}\in\mathbb{A}^{M\times N}$ represents the analog RF precoder where $\mathbb{A}=\big\{x|x\in\mathbb{C} \,\,\text{and}\,\,|x|=1\big\}$. Based on the model introduced in Section~\ref{Sec:Power}, to compensate the losses incurred in the RF network, $N$ amplifiers are needed in front of the power dividers, $MN$ amplifiers are needed in front of the phase shifters, and $MN$ amplifiers are needed in front of the power combiners. In total, we need $N(1+2M)\approx 2MN$ amplifiers when $M\gg 1$. Therefore, the total power consumption is obtained as 
\begin{IEEEeqnarray}{lll}\label{Eq:PowerFC}
	\Ptot =\Pbb+N\Prfc+2MN\Pamp+\Ptx/\rho_{\mathrm{pa}}.
\end{IEEEeqnarray}
Note that for large $M$, the loss caused by the power dividers and power combiners may exceed the maximum gain that the power amplifiers can provide without introducing non-linear distortions, see \cite[Fig.~7]{yan2018performance} for an example setup. Thereby, further amplification is needed within each stage of power division/combining. However, for simplicity, we neglect the additional power consumption in this paper.

\subsubsection{Partially-Connected MIMO Architectures} As can be seen from (\ref{Eq:PowerFC}), a huge challenge of the FC hybrid structure is scalability with respect to the number of antennas $M$. To address this issue, the PC hybrid MIMO structure has been proposed in the literature  \cite{gao2016energy}. The signal model for the PC architecture is identical to that in (\ref{Eq:Sig_Full}), i.e., $\mathbf{F} = \mathbf{R}\mathbf{B}$, with the difference that $\mathbf{R}$ is now a block-diagonal matrix
\begin{IEEEeqnarray}{lll}\label{Eq:Rpc} 
	\mathbf{R} = \begin{bmatrix}
		\mathbf{r}_{1} & \boldsymbol{0}_{r_1} &\cdots & \boldsymbol{0}_{r_1} \\
		\boldsymbol{0}_{r_2} & \mathbf{r}_{2}&\cdots & \boldsymbol{0}_{r_2}  \\
		\vdots &\vdots &\ddots &\vdots \\
		\boldsymbol{0}_{r_N} & \boldsymbol{0}_{r_N} &\cdots & \mathbf{r}_{N} 
	\end{bmatrix},
\end{IEEEeqnarray}
where $\mathbf{r}_n\in\mathbb{A}^{r_n\times 1}$ is the RF precoder vector which connects the output of the $n$-th RF chain to $r_n$ antennas, and $\boldsymbol{0}_{n}$ is a vector of length $n$ with all elements being equal to one. Note that $\sum_{n=1}^N r_n=M$ has to hold. In the simplest case, all RF chains are connected to the same number of antennas, i.e., $r_n=M/N,\,\,\forall n$, where we assume that $N$ is a divisor of $M$. Since the PC architecture does not require power combiners, only $N$ amplifiers are needed in front of the power dividers and $M$ amplifiers are needed in front of the phase shifters, i.e., there are $N+M\approx M$ amplifiers in total. Therefore, the total power consumption for the PC MIMO architecture is given by
\begin{IEEEeqnarray}{lll}\label{Eq:PowerPC}
	\Ptot =\Pbb+N\Prfc+M\Pamp+\Ptx/\rho_{\mathrm{pa}}.
\end{IEEEeqnarray}
 
 \begin{table*}
 	\label{Table:Comparison}
 	\caption{Comparison of Different MIMO Architectures, namely Fully Digital (FD), Fully Connected (FC), Partially Connected (PC), Reflect-Array (RA), and Transmit-Array (TA).    \vspace{-0.3cm}} 
 	\begin{center}
 		\scalebox{0.65}
 		{
 			\begin{tabular}{|| c | c | c | c ||}
 				\hline 
 				Architecture & Precoder $\mathbf{F}$ & Constraint & Total Power Consumption $\Ptot$ \\ \hline \hline
 				FD & $\mathbf{B}$ & $\mathbf{B}\in\mathbb{C}^{M\times Q}$ & $\Pbb+M\Prfc+\Ptx/\rho_{\mathrm{pa}}$\\ \hline   
 				FC    &  $\mathbf{R} \mathbf{B}$ & $\mathbf{B}\in\mathbb{C}^{N\times Q}$, $\mathbf{R}\in\mathbb{A}^{M\times N}$ & $\Pbb+N\Prfc+2MN\Pamp+\Ptx/\rho_{\mathrm{pa}}$\\ \hline   
 				PC    &  $\mathbf{R} \mathbf{B}$ & $\mathbf{B}\in\mathbb{C}^{N\times Q}$, $\mathbf{R}=\mathrm{diag}\{\mathbf{r}_1,\dots,\mathbf{r}_N\}, \mathbf{r}_n\in\mathbb{A}^{r_n\times 1}$, $\sum_{n=1}^N r_n=M$ & $\Pbb+N\Prfc+M\Pamp+\Ptx/\rho_{\mathrm{pa}}$\\ \hline   
 				RA \& TA    &  $c\mathbf{DTB}$ & $\mathbf{B}\in\mathbb{C}^{N\times Q}$, $\mathbf{T}\in\mathbb{C}^{M\times N}$ is a fixed matrix (see~(\ref{Eq:Tmaxtrix})), $\mathbf{D}=\mathrm{diag}\{d_1,\dots,d_M\}, d_m\in\mathbb{A}, c=\sqrt{\rho_{\mathrm{ary}}}\lambda/(4\pi)$ &$\Pbb+N\Prfc+\Ptx/(\rho_{\mathrm{rta}}\rho_{\mathrm{pa}})$\\ \hline   
 			\end{tabular}
 		} 
 	\end{center}\vspace{-0.3cm}
 \end{table*}
 
\subsection{Reflect-Array and Transmit-Array MIMO Architectures}\label{Sec:RA_TA}

In the following, we first model the constraints that the RA and TA architectures impose on the precoder.  Subsequently, we quantify the total power consumptions of these architectures.

\subsubsection{Constraints on the Hybrid Precoder} 
For the considered RA and TA architectures, we assume that each active feed antenna is connected to a dedicated RF chain, i.e., there are $N$ active antennas. Moreover, we assume that the passive array comprises $M$ antenna elements. To facilitate presentation, we characterize the positions of the passive antenna elements by $(r_{m,n},\phi_{m,n},\theta_{m,n})$ in $N$ different spherical coordinate systems corresponding to the locations of the active antennas such that each active antenna is the origin of one coordinate system and the $z$-axis is the direction of the main lobe of the antenna pattern. Note that the values of $(r_{m,n},\phi_{m,n},\theta_{m,n})$ depend on the specific configuration of the feed antenna and the array antennas. We further assume that the passive antennas have an omni-directional antenna pattern.

\begin{prop}\label{Prop:Precoder}
	Assuming  $r_{m,n}\gg \lambda$, the precoder $\mathbf{F}$ for the RA and TA antennas has the form
	\begin{IEEEeqnarray}{lll}\label{Eq:Precoder} 
		\mathbf{F} = \frac{\lambda\sqrt{\rho_{\mathrm{ary}}}}{4\pi}\mathbf{D}\mathbf{T}\mathbf{B},
	\end{IEEEeqnarray}
	where $\rho_{\mathrm{ary}}$ denotes the passive array power efficiency, $\mathbf{B}\in\mathbb{C}^{N\times Q}$ is the digital baseband precoder, $\mathbf{D}\in\mathbb{C}^{M\times M}$  is a diagonal matrix which controls the phase shifters and is given~by
	\begin{IEEEeqnarray}{lll}\label{Eq:Dmatrix} 
		\mathbf{D}=\mathrm{diag}\left(e^{j2\pi\beta_1},\dots,e^{j2\pi\beta_M}\right),
	\end{IEEEeqnarray}
	where $\beta_m\in[0,1]$, and $\mathbf{T}\in\mathbb{C}^{M\times N}$ is a fixed matrix which depends on the antenna configuration, namely $(r_{m,n},\phi_{m,n},\theta_{m,n}),\,\,\forall m,n$, and is given~by
	\begin{IEEEeqnarray}{lll} \label{Eq:Tmaxtrix}
		\mathbf{T}=\left[\sqrt{G(\theta_{m,n},\phi_{m,n})}\frac{e^{-j\frac{2\pi r_{m,n}}{\lambda}}}{r_{m,n}} \right]_{m,n}.
	\end{IEEEeqnarray}
\end{prop}
\begin{IEEEproof}
	The proof is given in Appendix~\ref{App:Prop_Precoder}.
\end{IEEEproof}

Proposition~\ref{Prop:Precoder} states that both RA and TA antennas have identical precoder structures, as given by (\ref{Eq:Precoder}), except that $\rho_{\mathrm{ary}}$ may assume different values for RA and TA antennas, see Section~\ref{Sec:Loss_RA_TA} for details. Therefore, the proposed precoder in Section~\ref{Sec:Precoder} can be applied to both RA and TA antennas.

\subsubsection{Power Consumption and Losses}\label{Sec:Loss_RA_TA}
We assume passive arrays for the RA and TA architectures, i.e., there is no signal amplification and $N_{\mathrm{amp}}=0$. Nevertheless, we have several power losses due to propagation over the air and other inefficiencies which are discussed in detail in the following: 

\textbf{Spillover loss:} Since the effective area of the array is finite,  some of the power radiated by the active antenna will not be captured by the passive antennas, resulting in a spillover loss \cite{pozar1997design}. We define the efficiency factor $\rho_{S}$ to take into account the spillover.

\textbf{Taper loss:} In general, the density of the received power differs across the passive antennas due to their corresponding different values of $G(\theta_{m,n},\phi_{m,n})$ and $r_{m,n}$. For single-stream transmission, it is well-known that the non-uniform power distribution across the passive antennas leads to a reduction of the achievable antenna gain and is referred to as taper loss \cite[Chapter~15]{balanis1982antenna}. For multiple-stream transmission, taper loss leads to a  reduction of the achievable rate. We define the efficiency factor $\rho_{T}$ to account for this loss. 

\textbf{Aperture loss:}  Ideally, for RA antennas, the total power captured by the aperture will be reflected. In practice, however, a certain fraction of the captured power may be absorbed by the RA. Similarly, for TA antennas, the aperture may not be able to fully forward the captured power and some of the power may be reflected in the backward direction or be absorbed by the TA. The aperture power efficiency is taken into account by introducing the efficiency factor $\rho_{A}$.

\textbf{Phase shifters:} Each phase shifter introduces a certain loss which is captured by the efficiency factor $\rho_{P}$. For TA antenna, the received signal passes through the phase shifter once before being forwarded whereas for RA antennas, the signal passes through the phase shifter twice before being reflected. Hence, the overall phase shifter efficiency factors for RA and TA antennas are $\rho^2_{P}$ and $\rho_{P}$, respectively.


Note that the effects of the spillover and taper losses are included in  matrix $\mathbf{T}$ in (\ref{Eq:Tmaxtrix}). Therefore, the array efficiency for RA and TA antennas is obtained as $\rho_{\mathrm{ary}}=\rho^2_{P}\rho_{A}$ and $\rho_{\mathrm{ary}}=\rho_{P}\rho_{A}$, respectively, which accounts for the combined effects of the aperture and phase shifter losses. The transmit power is $\Ex\{\mathbf{x}\Herm\mathbf{x}\}=\Ptx\|\mathbf{F}\|_F^2=\Ptx$ where $\mathbf{x}=\sqrt{\Ptx}\mathbf{Fs}$ and $\|\mathbf{F}\|_F=1$ whereas the  power radiated by the active antennas is $\Prd=\Ex\{\bar{\mathbf{x}}\Herm\bar{\mathbf{x}}\}=\Ptx\|\mathbf{B}\|_F^2$ where $\bar{\mathbf{x}}=\sqrt{\Ptx}\mathbf{Bs}$ is the signal transmitted over active antennas. In fact, due to the aforementioned power losses\footnote{Note that taper loss reduces the achievable rate but does not constitute a power loss.}, i.e., $\rho_{S}$,  $\rho_{P}$, and $\rho_{A}$, the power radiated by the active antennas $\Prd$ is not identical to the power radiated by the passive antennas $\Ptx$. Therefore, the total power loss is obtained as
\begin{IEEEeqnarray}{lll}\label{Eq:PowerRA_TA}
	\Ptot &=\Pbb+N\Prfc+\Ptx\|\mathbf{B}\|_F^2/\rho_{\mathrm{pa}} \nonumber \\
	&\approx\Pbb+N\Prfc+\Ptx/(\rho_{\mathrm{rta}}\rho_{\mathrm{pa}}),
\end{IEEEeqnarray}
where $\rho_{\mathrm{rta}}=\rho_{S}\rho_{\mathrm{ary}}$. As can be seen from (\ref{Eq:PowerRA_TA}), unlike for conventional MIMO architectures, the total power consumption of the RA and TA antennas does not explicitly change with increasing number of passive antennas $M$ which may lead to an improved energy-efficiency and scalability. The constraints imposed on the precoder and the total power consumption of the different massive MIMO architectures discussed in this paper are summarized in Table~I.

\section{Precoding Design}\label{Sec:Precoder}

In this section, we propose an efficient linear precoder design for RA and TA antennas exploiting the sparsity of the mmWave channel. Ideally, we would determine the optimal precoder which maximizes the achievable rate, denoted by $R$, based on
\begin{IEEEeqnarray}{rll}\label{Eq:RateMax} 
\underset{\mathbf{F}\in\mathcal{F}}{\mathrm{maximize}} \,\,&R= \log_2\left(\Big|\mathbf{I}_{J}+\frac{\Ptx}{\sigma^2}\mathbf{H}\mathbf{F}\mathbf{F}\Herm\mathbf{H}\Herm\Big|\right) \nonumber \\
\mathrm{C1}\hspace{-1mm}:\,\, &\|\mathbf{F}\|_F^2\leq 1,
\end{IEEEeqnarray}
where $\mathrm{C1}$ enforces the transmit power constraint and $\mathcal{F}$ is the set of feasible precoders which depends on the adopted MIMO architecture. For instance,  for FC hybrid MIMO, we have $\mathcal{F}=\{\mathbf{F}=\mathbf{RB}|\mathbf{B}\in\mathbb{C}^{N\times Q} \,\,\text{and}\,\, \mathbf{R}\in\mathbb{A}^{M\times N}\}$. Unfortunately, problem (\ref{Eq:RateMax}) is not tractable for hybrid MIMO architectures (including the considered RA and TA) since set $\mathcal{F}$ is not convex due to modulo-one constraint on the elements of the analog precoder, cf. (\ref{Eq:Sig_Full}), (\ref{Eq:Rpc}), and (\ref{Eq:Precoder}). Let $\mathbf{F}^{\mathrm{opt}}$ denote the optimal unconstrained precoder for the FD MIMO architecture, i.e., $\mathcal{F}=\mathbb{C}^{M\times Q}$. Instead of (\ref{Eq:RateMax}), minimization of $\|\mathbf{F}^{\mathrm{opt}}-\mathbf{F}\|_F$ is commonly adopted in the literature  as design criterion for constrained hybrid precoders \cite{el2014spatially,ahmed2018survey,zhou2018hardware,lin2016energy}. Therefore, we consider the following optimization problem for the RA and TA hybrid MIMO architectures 
\begin{IEEEeqnarray}{rll} \label{Eq:ProbMin}
\underset{\mathbf{B}\in\mathcal{B},\mathbf{D}\in\mathcal{D}}{\mathrm{minimize}} \,\, &\big\|\mathbf{F}^{\mathrm{opt}} - c \mathbf{D}\mathbf{T}\mathbf{B} \big\|_F^2 \nonumber \\
 \mathrm{C1}\hspace{-1mm}:\,\, &\|c\mathbf{DTB}\|_F^2\leq 1,
\end{IEEEeqnarray}
where $c=\frac{\lambda\sqrt{\rho_{\mathrm{ary}}}}{4\pi}$, $\mathcal{B}=\big\{\mathbf{B}\in\mathbb{C}^{N\times Q}\big\}$, and $\mathcal{D}=\big\{\mathbf{D}\in\mathbb{C}^{M\times M} | \mathbf{D}=\mathrm{diag}\{d_1,\dots,d_M\}, d_m\in\mathbb{A}\big\}$. Note that (\ref{Eq:ProbMin}) is still non-convex due to multiplication of $\mathbf{D}$ and $\mathbf{B}$  as well as the non-convexity of set $\mathcal{D}$. Nevertheless, (\ref{Eq:ProbMin}) allows us to design an efficient suboptimal solution in the following.

\subsubsection{Rationale Behind the Proposed Precoder} Let $\mathbf{H}=\mathbf{U}\boldsymbol{\Sigma}\mathbf{V}\Herm$ denote the singular value decomposition (SVD) of  channel matrix $\mathbf{H}$, where $\mathbf{U}$ and $\mathbf{V}$ are unitary matrices containing the left and right singular vectors, respectively, and $\boldsymbol{\Sigma}$ is a diagonal matrix containing the singular values. The optimal unconstrained precoder is $\mathbf{F}^{\mathrm{opt}}= [\alpha_1\mathbf{v}_1,\alpha_2\mathbf{v}_2,\dots,\alpha_Q\mathbf{v}_Q]$ where $\mathbf{v}_q$ is the right singular vector corresponding to the $q$-th largest singular value and $\alpha_q$ is the power allocation factor obtained via the water filling algorithm \cite{biglieri2007mimo}. For the spatially sparse channel model introduced in (\ref{Eq:Channel}),  $\mathcal{H}_t=\big\{\mathbf{h}_t(\theta_l^t,\phi_l^t),\forall l=1,\dots,L\big\}$ forms a vector space  for the rows of $\mathbf{H}$. In addition, since $L\ll M$ and $(\theta_l^t,\phi_l^t)$ is taken from a continuous distribution, the elements of $\mathcal{H}_t$ are with probability one linearly independent. Thereby, the columns of $\mathbf{F}^{\mathrm{opt}}$ can be written as a linear combination of the transmit array response, i.e., $\mathbf{v}_q=\sum_{l}c_{l,q}\mathbf{h}_t(\theta_l^t,\phi_l^t)$ where $c_{l,q}$ are the corresponding coefficients \cite{el2014spatially}. More compactly, $\mathbf{F}^{\mathrm{opt}}$ can be rewritten as
\begin{IEEEeqnarray}{lll} \label{Eq:OptimalF}
\mathbf{F}^{\mathrm{opt}}=\mathbf{H}_t\mathbf{C},
\end{IEEEeqnarray}
where $\mathbf{H}_t=[\mathbf{h}_t(\theta_1^t,\phi_1^t),\dots,\mathbf{h}_t(\theta_L^t,\phi_L^t)]\in\mathbb{C}^{M\times L}$ and $\mathbf{C}\in\mathbb{C}^{L\times Q}$ contains the coefficients $\alpha_qc_{l,q}$. The similarity of the structure of the optimal precoder in (\ref{Eq:OptimalF}) and the hybrid precoder  $\mathbf{F}=\mathbf{R}\mathbf{B}$ has motivated researchers to use the channel response vectors $\mathbf{h}_t(\theta_l^t,\phi_l^t)$ for the columns of $\mathbf{R}$. Since $\mathbf{R}$ has $N$ columns (i.e., $N$ RF chains), the problem in  (\ref{Eq:ProbMin}) can be approximated as choosing the best $N$ columns of $\mathbf{H}_t$ to approximate $\mathbf{F}^{\mathrm{opt}}$ \cite{gao2018low,ahmed2018survey,el2014spatially}. Unfortunately, this concept is not directly applicable to the precoder in (\ref{Eq:Precoder}) because of its different structure. Hence, we rewrite  (\ref{Eq:OptimalF}) in a more useful form. Let us divide the index set of the passive antennas $\{1,\dots,M\}$ into $N$ mutually exclusive sets $\mathcal{M}_{n},\,\,n=1,\dots,N$. Thereby, (\ref{Eq:OptimalF}) can be rewritten as
\begin{IEEEeqnarray}{lll} \label{Eq:OptimalFreform}
	\mathbf{F}^{\mathrm{opt}}=\sum_{n=1}^{N} \mathbf{H}_t^{\mathcal{M}_{n}}\mathbf{C},
\end{IEEEeqnarray}
where $\mathbf{H}_t^{\mathcal{M}_{n}}=\mathbf{I}_{\mathcal{M}_{n}}\mathbf{H}_t\in\mathbb{C}^{M\times L}$ and $\mathbf{I}_{\mathcal{M}_{n}}\in\{0,1\}^{M\times M}$ is a diagonal matrix whose $m$-th diagonal entry is one if $m\in\mathcal{M}_n$ and zero otherwise. Now, let us rewrite the precoder in (\ref{Eq:Precoder}) as
\begin{IEEEeqnarray}{lll} \label{Eq:Freform}
	\mathbf{F}=c\sum_{n=1}^{N} \mathbf{D}^{\mathcal{M}_{n}}\mathbf{T}^{\mathcal{M}_{n}}\mathbf{B},
\end{IEEEeqnarray}
where $\mathbf{D}^{\mathcal{M}_{n}}=\mathbf{I}_{\mathcal{M}_{n}}\mathbf{D}\in\mathbb{A}^{M\times N}$ and $\mathbf{T}^{\mathcal{M}_{n}}=\mathbf{I}_{\mathcal{M}_{n}}\mathbf{T}\in\mathbb{C}^{M\times N}$. Comparing (\ref{Eq:OptimalFreform}) and (\ref{Eq:Freform}) motivates us to choose $\mathbf{D}^{\mathcal{M}_{n}}$ such that $\mathbf{D}^{\mathcal{M}_{n}}\mathbf{T}^{\mathcal{M}_{\bar{n}}}$ becomes similar to $\mathbf{H}_t^{\mathcal{M}_{n}}$. To do this, we have to address the following two challenges. First, since $\mathbf{D}^{\mathcal{M}_{n}}$ has only $M/N$ non-zero elements and $\mathbf{H}_t^{\mathcal{M}_{n}}$ has $ML/N$ non-zero elements, $\mathbf{H}_t^{\mathcal{M}_{n}}$ cannot be fully reconstructed by $\mathbf{D}^{\mathcal{M}_{n}}\mathbf{T}^{\mathcal{M}_{n}}$. Hereby, we choose to reconstruct only one column of $\mathbf{H}_t^{\mathcal{M}_{n}}$ by $\mathbf{D}^{\mathcal{M}_{n}}\mathbf{T}^{\mathcal{M}_{n}}$. The unmatched columns of $\mathbf{D}^{\mathcal{M}_{n}}\mathbf{T}^{\mathcal{M}_{n}}$ are treated as interference. Fortunately, for large $M$, the interference approaches zero due to channel hardening. Second, we have to choose which column of $\mathbf{H}_t^{\mathcal{M}_{n}}$ to reconstruct. In this paper, we employ  OMP to choose the best $N$ columns of $\mathbf{H}_t^{\mathcal{M}_{n}}$. Based on these insights, we present the proposed precoder in the following.

\begin{table*}
	\label{Table:Parameter}
	\caption{Default Values of System Parameter \cite{yan2018performance,ribeiro2018energy,lin2016energy,lau2012reconfigurable,di2015reconfigurable}.\vspace{-0.1cm}} 
	\begin{center}
		\scalebox{0.55}
		{
			\begin{tabular}{|| c | c | c | c | c | c | c | c | c | c | c | c | c | c | c | c | c | c | c | c | c | c | c | c  ||}
				  \hline\vspace{-0.3cm}
				   & & & & & &  & & & & & & & & & & & & &\\ 
				Parameter & $\ell$ & $\eta$ & $\bar{\theta}_l^t,\bar{\theta}_l^r$ & $\bar{\phi}_l^t,\bar{\phi}_l^r$ & $L$ & $N_0$ & $N_F$ & $W$ 
				& $\lambda$ & $d$ & $R_r$ & $R_d$ & $\kappa$ & $\rho_P$ & $\rho_A$ & $\Pbb$ & $\Prfc$ & $\Pamp$ & $\rho_{\mathrm{amp}}$ \\ \hline\hline
				\vspace{-0.35cm}
				& & & & & &  & & & & & & & & & & & & &\\
				 Value & $100$~m  & $2$ & $\pi/3$ & $2\pi/3$ & $8$ & $-174$~dBm/Hz & $6$ dB & $100$ MHz & $5$~mm  & $\lambda/2$ & FI:$2d$, PI:$\frac{d\sqrt{2M}}{4}$ 
				& FI:$\frac{d\sqrt{M}}{\sqrt{\pi}}$, PI:$\frac{d\sqrt{M}}{\sqrt{4\pi}}$ & $6$ & $-2$ dB & RA: $-0.5$ dB, TA: $-1.5$ dB &  $200$ mW & $120$ mW & $40$ mW & $0.3$ \\ \hline 
			\end{tabular}
		} 
	\end{center}
\end{table*}

\subsubsection{Proposed Precoder} Let us fix sets $\mathcal{M}_{n},\,\,n=1,\dots,N$ a priori.  The proposed precoder employs $N$ iterations where in each iteration, the following three steps are performed:

\textbf{Step~1--Choosing the Next Dimension:}  Let $\mathbf{F}_i^{\mathrm{res}}=\mathbf{F}^{\mathrm{opt}}-c\sum_{n=1}^{i} \mathbf{D}^{\mathcal{M}_{n}}\mathbf{T}^{\mathcal{M}_{n}}\mathbf{B}_i$ denote the residual precoder in iteration $i$ where $\mathbf{B}_{i}$ is the baseband precoder designed in iteration~$i$. In each iteration, we project the residual matrix from the previous iteration on the space defined by $\mathbf{H}_t$ and find the direction $l^*$ that has the maximum projected value. This can be mathematically formulated as
\begin{IEEEeqnarray}{lll} \label{Eq:Angle}
l_i^*=\mathrm{argmax}_{l=1,\dots,L}\,\,(\boldsymbol{\Psi}_i\boldsymbol{\Psi}_i\Herm)_{l,l},
\end{IEEEeqnarray}
where $\boldsymbol{\Psi}_i=\mathbf{H}_t\Herm \mathbf{F}_{i-1}^{\mathrm{res}}$.
              
\textbf{Step~2--Computation of $\mathbf{D}^{\mathcal{M}_{n}}$:} $\mathbf{D}^{\mathcal{M}_{n}}$ is initialized to the zero matrix $\boldsymbol{0}_{M\times M}$. We obtain the diagonal elements of $\mathbf{D}^{\mathcal{M}_{n}}$ corresponding to the indices in set $\mathcal{M}_n$ as
\begin{IEEEeqnarray}{lll} \label{Eq:Dopt}
\mathbf{D}^{\mathcal{M}_{n}}_{m,m} = \exp\left(j\left[\angle (\mathbf{H}_t^{\mathcal{M}_{n}})_{m,l_i^*} - \angle \mathbf{T}_{m,n}\right]\right),\,\,\forall m\in\mathcal{M}_n. \quad
\end{IEEEeqnarray}
In other words, the passive antennas $m\in\mathcal{M}_n$ create a coherent wave plane in direction $l_i^*$ for the signal illuminated by the $n$-th active antenna. 

\textbf{Step~3--Computation of $\mathbf{B}_i$:} By defining $\mathbf{W}_i=c\sum_{n=1}^{i} \mathbf{D}^{\mathcal{M}_{n}}\mathbf{T}^{\mathcal{M}_{n}}$, we can formulate the following optimization problem for $\mathbf{B}_i$ 
\begin{IEEEeqnarray}{lll} \label{Eq:ProbB}
\mathbf{B}_i = &\underset{\mathbf{B}\in\mathcal{B}}{\mathrm{argmin}}\,\, \big\|\mathbf{F}^{\mathrm{opt}} -  \mathbf{W}_i\mathbf{B} \big\|_F^2,
\nonumber \\
&\mathrm{C1}:\,\, \|\mathbf{W}_i\mathbf{B}\|_F^2\leq \Ptx,
\end{IEEEeqnarray}
 which has the following well-known normalized least square solution \cite{el2014spatially} 
\begin{IEEEeqnarray}{lll} \label{Eq:SolB}
\mathbf{B}_i  = \frac{(\mathbf{W}_i\Herm\mathbf{W}_i)^{-1}\mathbf{W}_i\Herm\mathbf{F}^{\mathrm{opt}}}
{\|\mathbf{W}_i(\mathbf{W}_i\Herm\mathbf{W}_i)^{-1}\mathbf{W}_i\Herm\mathbf{F}^{\mathrm{opt}}\|_F}.
\end{IEEEeqnarray}
Note that $\mathbf{B}_i$ effectively eliminates the interference between the data streams.

Algorithm~\ref{Alg:PrecoderSparse} summarizes the above main steps for the proposed OMP-based precoder design.

\begin{algorithm}[t]
\caption{OMP-based Precoder Design}
 \begin{algorithmic}[1]\label{Alg:PrecoderSparse}
 \STATE \textbf{initialize:} $\mathbf{F}_0^{\mathrm{res}}=\mathbf{F}^{\mathrm{opt}}$ and $\mathbf{D}^{\mathcal{M}_{n}}=\boldsymbol{0}_{M\times M},\,\,\forall n$.  
      \FOR{$i=1,\dots,N$}
              \STATE $l_i^*=\mathrm{argmax}_{l=1,\dots,L}\,\,(\boldsymbol{\Psi}_i\boldsymbol{\Psi}_i\Herm)_{l,l}$ 
              where $\boldsymbol{\Psi}_i=\mathbf{H}_t\Herm \mathbf{F}_{i-1}^{\mathrm{res}}$.
              \STATE Update $\mathbf{D}^{\mathcal{M}_{n}}$ using (\ref{Eq:Dopt}).
              \STATE Update $\mathbf{B}_i$ using (\ref{Eq:SolB}) for $\mathbf{W}_i=c\sum_{n=1}^{i} \mathbf{D}^{\mathcal{M}_{n}}\mathbf{T}^{\mathcal{M}_{n}}$.
              \STATE Update $\mathbf{F}_i^{\mathrm{res}}=\mathbf{F}^{\mathrm{opt}}-c\sum_{n=1}^{i} \mathbf{D}^{\mathcal{M}_{n}}\mathbf{T}^{\mathcal{M}_{n}}\mathbf{B}_i$.
      \ENDFOR
      \STATE Return $\mathbf{D}=\sum_{n=1}^{N} \mathbf{D}^{\mathcal{M}_{n}}$ and $\mathbf{B}=\mathbf{B}_N$.
  \end{algorithmic}
\end{algorithm}

\section{Simulation Results} 
In this section, we first describe the considered simulation setup and introduce the adopted benchmark schemes. Subsequently, we compare  the performances of the considered mmWave massive MIMO architectures.

\subsection{Simulation Setup}
We generate the channel matrices according to (\ref{Eq:Channel}). Thereby, we assume that the angles $\theta_l^t$, $\theta_l^r$,  $\phi_l^t$, and $\phi_l^r$ are  uniformly distributed RVs in the intervals $[0,\bar{\theta}_l^t]$, $[0,\bar{\theta}_l^r]$, $[0,\bar{\phi}_l^t]$, and $[0,\bar{\phi}_l^r]$, respectively, and $\bar{\theta}_l^t$ and $\bar{\theta}_l^r$ ($\bar{\phi}_l^t$ and $\bar{\phi}_l^r$) are the elevation (azimuth) coverage angles of the transmitter and receiver antennas, respectively. Moreover, we use a square uniform planar array in (\ref{Eq:ULA}), i.e., a $\sqrt{M}\times\sqrt{M}$ planar array. The channel coefficient for each effective path is modeled as $h_l=\sqrt{\bar{h}_l}\tilde{h}_l$ where $\bar{h}_l$ and $\tilde{h}_l$ are the path loss and the random fading components, respectively, and are given by
\begin{IEEEeqnarray}{lll}\label{Eq:PathLossRayleigh}
	\bar{h}_l = \left(\frac{\lambda}{4\pi \ell}\right)^{\eta} 
	\quad \text{and} \quad
	\tilde{h}_l = \mathcal{CN}(0,1),
\end{IEEEeqnarray}
respectively. In (\ref{Eq:PathLossRayleigh}), $\ell$ denotes the distance between the transmitter and the receiver and $\eta$ represents the path-loss exponent. The noise power at the receiver is given by $\sigma^2=WN_0N_F$ where $W$ is the bandwidth, $N_0$ represents the noise power spectral density, and $N_F$ denotes the noise figure. We arrange the active antennas with respect to the array of passive antennas as follows. All active antennas have distance $R_d$ from the passive array and are located on a ring of radius $R_r$. The line that connects the center of the ring to the center of the plane is perpendicular to the array plane. Moreover, we adopt the following simple class of axisymmetric feed antenna patterns, which is widely used in the antenna community  \cite{balanis1982antenna,pozar1997design},
\begin{IEEEeqnarray}{lll} 
G(\theta,\phi) = \begin{cases}
2(\kappa+1)\cos^\kappa(\theta),\quad & \mathrm{if}\,\,0\leq \theta \leq \frac{\pi}{2} \\
0,\quad & \mathrm{if}\,\, \frac{\pi}{2}< \theta \leq \pi , \\
\end{cases}
\end{IEEEeqnarray}
where $\kappa\geq 2$ is a number and normalization factor $2(\kappa+1)$ ensures that $\int_{\Omega} \frac{1}{4\pi} G(\phi,\theta)\mathrm{d}\Omega=1$ holds where $\mathrm{d}\Omega=\sin(\theta)\mathrm{d}\theta\mathrm{d}\phi$ \cite{balanis1982antenna}. The default values of the system parameters are provided in Table~II. The results shown in this section are averaged over $10^3$ random realization of the channel matrix.

\begin{figure*}
	\begin{minipage}{0.32\linewidth}
		\centering
		\resizebox{1\linewidth}{!}{\psfragfig{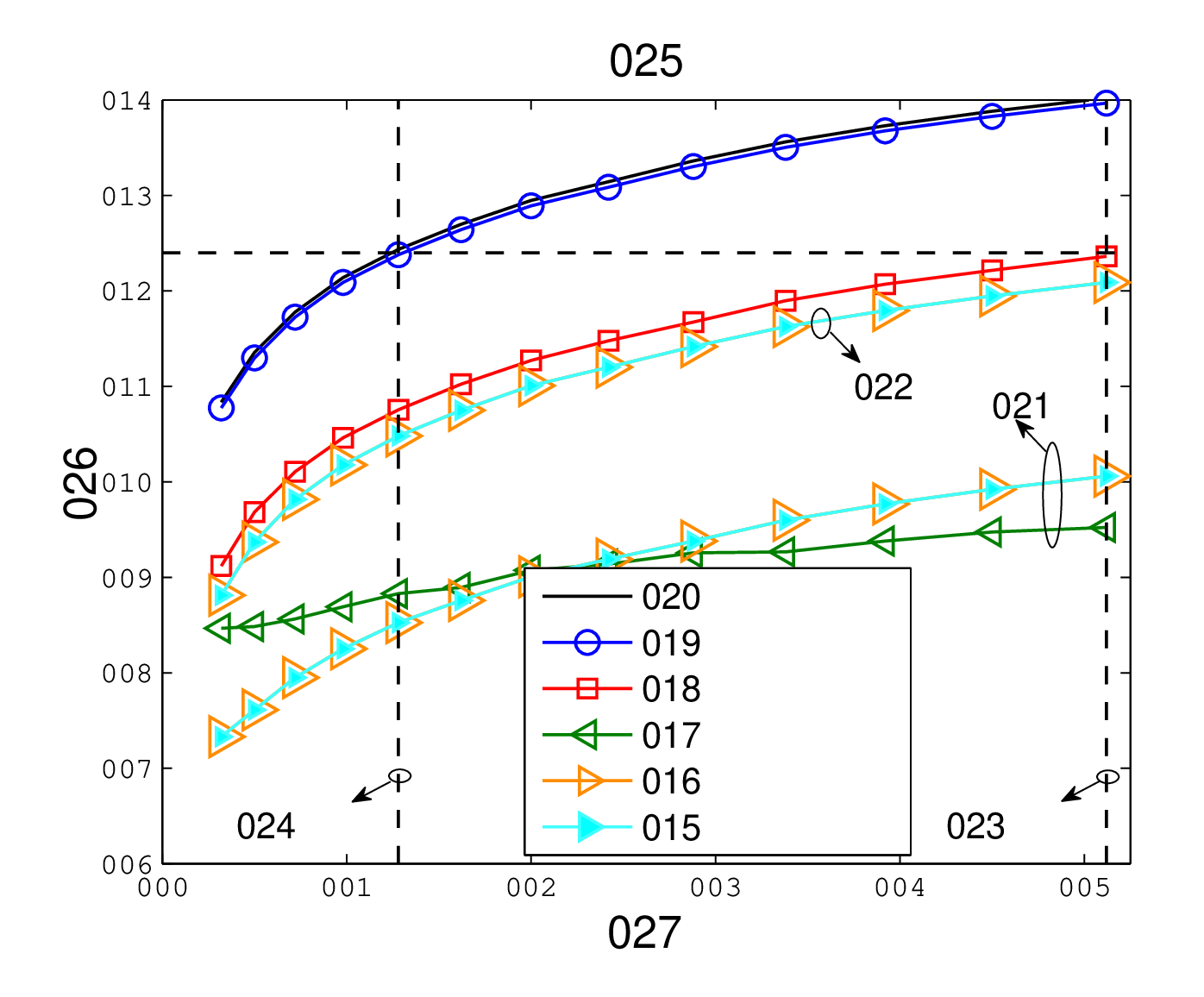}}
	\end{minipage}
	\begin{minipage}{0.32\linewidth}
		\centering
		\resizebox{1\linewidth}{!}{\psfragfig{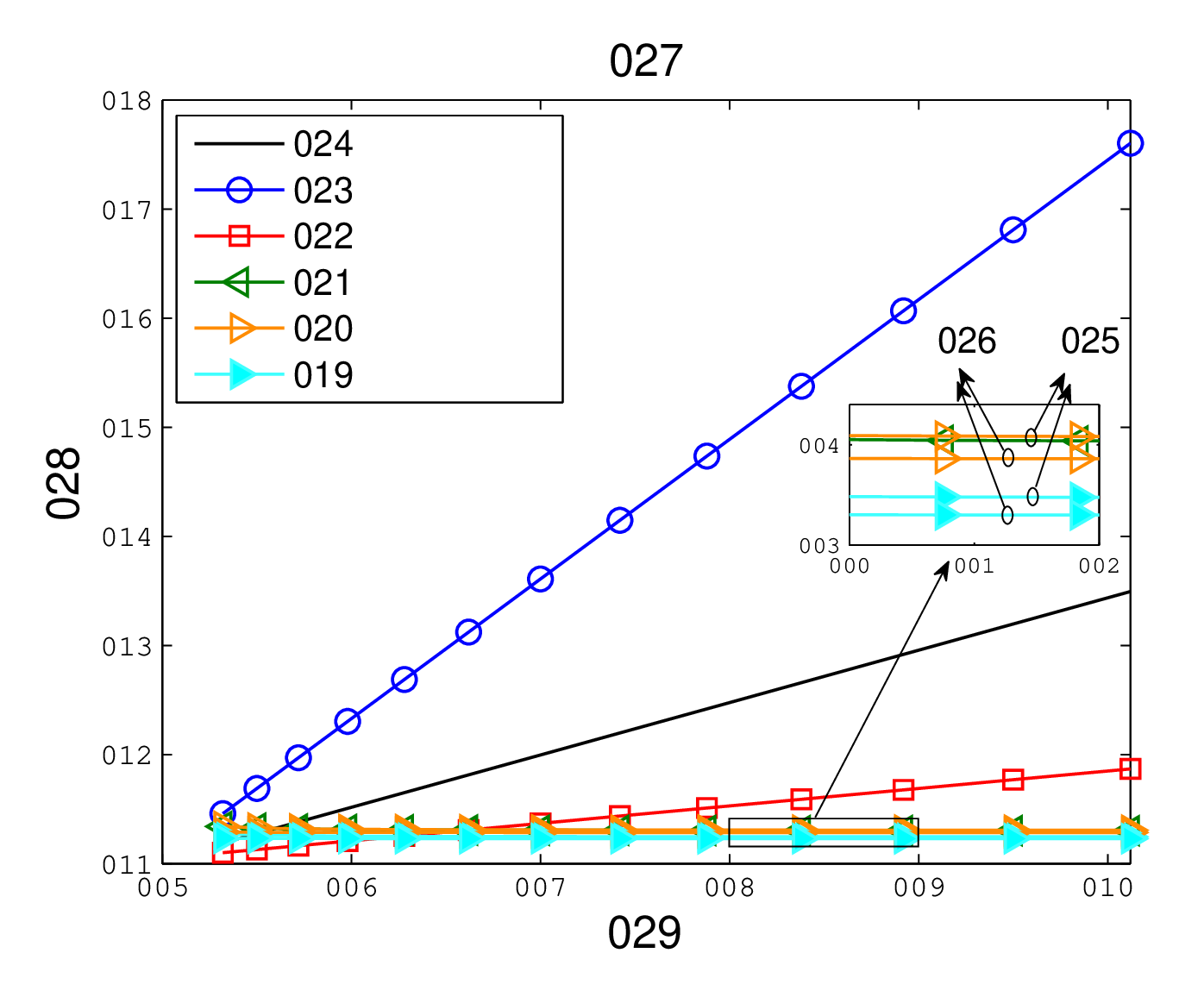}}
	\end{minipage}
	\begin{minipage}{0.32\linewidth}
		\centering
		\resizebox{1\linewidth}{!}{\psfragfig{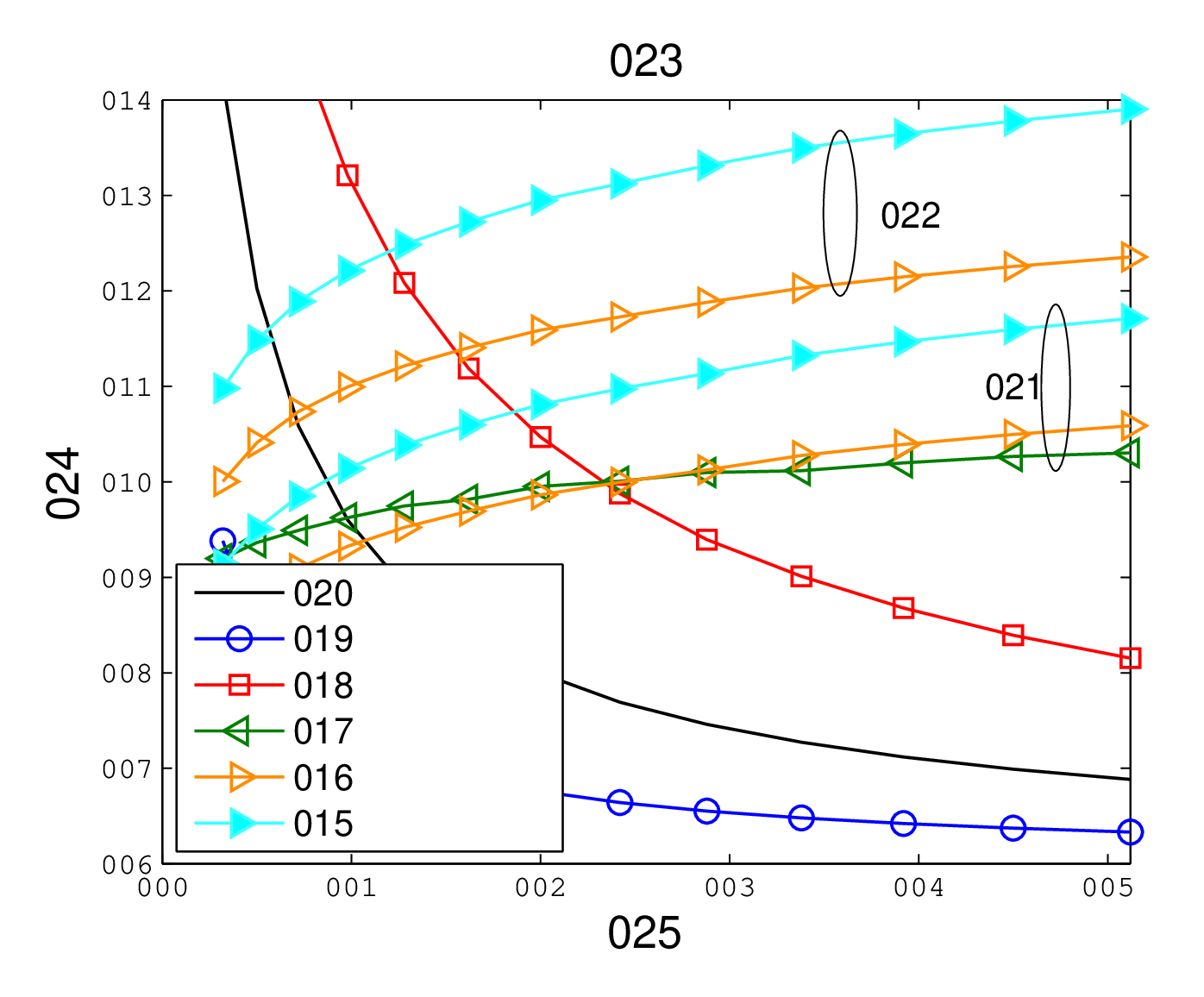}}
	\end{minipage}
	\caption{a) Spectral efficiency (bits/s/Hz), b) total consumed power $\Ptot$ (Watt), and c) energy efficiency (MBits/Joule) versus number of transmit antennas $M$ for $N=4$, $Q=4$, $J=64$, and $\Ptx=500$ mWatt. }
	\label{Fig:RatePowerEE_M}
\end{figure*}

\subsection{Benchmark Schemes}
For the FD MIMO architecture, we consider the optimal unconstrained precoder obtained from the SVD of the channel and water filling power allocation. For the FC hybrid architecture, we considered the spatially-sparse precoder introduced in \cite{el2014spatially}. We note that the precoder for the PC architecture can be rewritten as $\mathbf{F}=\mathbf{D}\tilde{\mathbf{T}}\mathbf{D}$ where $\tilde{\mathbf{T}}$ is a fixed matrix whose element in the $m$-th row and $n$-th column is one if the $m$-th antenna is connected to the $n$-th RF chain and zero otherwise. Therefore, we can apply the proposed precoder design also to the PC architecture. Finally,  we also use the AO-based precoder recently proposed for RA in \cite{zhou2018hardware}  as a benchmark. Note that in \cite{zhou2018hardware}, each active antenna illuminates the full passive array, referred to as full illumination (FI). However, for the proposed precoder, only a part of the passive array is responsible for reflection/transmission of the signal received from a given active antenna. Therefore, in addition to FI, we also consider the case where each active antenna mostly illuminates  the subset of  passive antenna elements allocated to it, referred to as partial illumination (PI)\footnote{We do not show the results for the precoder in \cite{zhou2018hardware} for PI since this precoder was not designed for PI and, as a result, has a poor performance in this case. Hence, the comparison would not be fair.}. This is achieved by proper configuration of the positions of active antennas with respect to the passive array via $R_r$ and $R_d$, see Table~II. 


\subsection{Performance Evaluation}

In Fig.~\ref{Fig:RatePowerEE_M}, we show a) the spectral efficiency $R$ (bits/s/Hz) given in (\ref{Eq:RateMax}), b) the total consumed power $\Ptot$ (Watt), and c) the energy efficiency, defined as $WR/\Ptot$, (Bits/Joule) versus the number of transmit antennas $M$ for $N=4$, $Q=4$, $J=64$, and $\Ptx=10$ Watt. 
As can be seen from Fig.~\ref{Fig:RatePowerEE_M} a), the FC hybrid architecture can closely approach the spectral efficiency of the FD architecture. As expected, PC hybrid MIMO  has a lower spectral efficiency compared to FC hybrid MIMO  due to the fewer degrees of freedom of PC MIMO for beamforming as $MN$ and $M$ phase shifters are used in the FC and PC architectures, respectively. Although the RA and TA  architectures have $M$ phase shifters, too, they achieve a lower spectral efficiency compared to the PC architecture since the superposition of the signals of the different active antennas occurs over the air and cannot be fully controlled. This creates an unintended interference between the signals transmitted from different active antennas. Nevertheless, this interference is considerably reduced for PI compared to FI which leads to a considerable improvement in spectral efficiency. Finally, we observe from Fig.~\ref{Fig:RatePowerEE_M} a) that under FI, the proposed OMP-based precoder outperforms the AO-based precoder in \cite{zhou2018hardware} for large $M$. This might be attributed to the fact that for large $M$, the iterative AO-based  algorithm in \cite{zhou2018hardware} is more prune to getting trapped in a local optimum which is avoided by the proposed OMP-based precoder which efficiently exploits the sparsity of the mmWave channel. On the other hand,  the proposed precoder with PI outperforms the precoder in \cite{zhou2018hardware} for the entire considered range of $M$ due to the reduction of interference between the signals emitted by different active antennas for partial illumination. Recall that RA and TA antennas have identical precoder structures given in (\ref{Eq:Precoder}) but different values of $\rho_{\mathrm{ary}}$ which affects their total power consumptions, cf. (\ref{Eq:PowerRA_TA}). This is the reason why in Fig.~\ref{Fig:RatePowerEE_M} a), RA and TA antennas yield identical spectral efficiency. 

The main advantage of the RA and TA architectures is their scalability in terms of the number of antennas $M$ which is evident from Figs.~\ref{Fig:RatePowerEE_M} b) and c). In fact, for PI, RA and TA MIMO achieve similar performance as FD and FC MIMO if they are equipped with $N$ times more antennas, e.g., in Fig.~\ref{Fig:RatePowerEE_M} a), FD and FC MIMO with $M=256$ antennas and RA and TA MIMO with $M=1024$ antennas (under PI) achieve the same spectral efficiency of $64$ bits/s/Hz. However, from  Fig.~\ref{Fig:RatePowerEE_M} b), we observe that the total transmit power of the conventional FD, FC, and PC architectures significantly increases as $M$ increases which makes their implementation quite costly or even infeasible. On the other hand, the total power consumption of the RA and TA architectures stays almost the same as $M$ increases. As a result, we observe in Fig.~\ref{Fig:RatePowerEE_M} c) that the energy efficiency of the conventional  FD, FC, and PC architectures decreases as  $M$ increases whereas the energy efficiency of the proposed RA and TA architectures increases. From Figs.~\ref{Fig:RatePowerEE_M} b), we observe that PI yields a lower power consumption than FI since each active antenna more efficiently uses its transmit power and illuminates mostly the part of passive array that is responsible for reflection/transmission of its signal. This leads to a higher energy efficiency of PI compared to FI in Fig.~\ref{Fig:RatePowerEE_M} c), too. From Figs.~\ref{Fig:RatePowerEE_M} b) and c), we observe that TA antennas have higher energy efficiency and lower power consumption compared to RA antennas which is due to higher array efficiency factor, i.e., $[\rho_{\mathrm{ary}}]_{\text{dB}}=2[\rho_P]_{\text{dB}}+[\rho_A]_{\text{dB}}=-4.5$~dB and $[\rho_{\mathrm{ary}}]_{\text{dB}}=[\rho_P]_{\text{dB}}+[\rho_A]_{\text{dB}}=-3.5$~ dB for RA and TA, respectively, cf. Table~II.  

\section{Conclusions}
In this paper, we considered hybrid RA and TA MIMO architectures, analyzed their precoder structure and consumed power, and compared them to conventional fully-digital and hybrid MIMO architectures. Moreover, we designed a precoder based on OMP which efficiently exploits the sparsity of mmWave channels. Our simulation results revealed that unlike conventional MIMO architectures, RA and TA MIMO architectures are highly energy efficient and fully scalable in terms of the number of transmit antennas.

\appendices
\section{Proof of Proposition~\ref{Prop:Precoder}}\label{App:Prop_Precoder}
Let us define $\bar{\mathbf{x}}=[\bar{x}_1,\dots,\bar{x}_N]\Trans\in\mathbb{C}^{N\times 1}$ and $\bar{\mathbf{y}}=[\bar{y}_1,\dots,\bar{y}_M]\Trans\in\mathbb{C}^{M\times 1}$ where $\bar{x}_n$ and $\bar{y}_m$ denote the signal transmitted by the $n$-th active antenna and the signal received at the $m$-th passive antenna, respectively. As can be seen from Fig.~\ref{Fig:SysMod}, the data stream vector $\mathbf{s}$ is multiplied by the baseband precoder $\mathbf{B}$, fed to the RF chains, and then transmitted over the active antennas/illuminators, i.e., $\bar{\mathbf{x}}=\sqrt{\Ptx}\mathbf{B}\mathbf{s}$. Let us assume $r_{m,n}\gg \lambda$ such that the passive antennas are in the far field with respect to the active antennas. Thereby,  the signal that is received at the $m$-th passive antenna, $\bar{y}_{m}$, is obtained as \cite{di2015reconfigurable}
\begin{IEEEeqnarray}{lll} 
\bar{y}_{m} = \sum_{n=1}^N\sqrt{\Ptx G(\theta_{m,n},\phi_{m,n})}\frac{\lambda}{4\pi r_{m,n}}e^{-j\frac{2\pi r_{m,n}}{\lambda}} \bar{x}_n.
\end{IEEEeqnarray}
Defining matrix $\mathbf{T}$ in (\ref{Eq:Tmaxtrix}), we obtain $\bar{\mathbf{y}}=\frac{\lambda}{4\pi}\mathbf{T}\bar{\mathbf{x}}$. At the passive antenna array, the received signal at the $m$-th antenna is delayed by phase $2\pi\beta_m$ and reflected/transmitted. Defining $\mathbf{D}$ in (\ref{Eq:Dmatrix}), we obtain $\mathbf{x}=\mathbf{D}\bar{\mathbf{y}}$. Until now, we did not include the signal attenuation due the aperture efficiency and phase shifter efficiency which is captured by the efficiency factor $\rho_{\mathrm{ary}}$, see Section~\ref{Sec:Loss_RA_TA}. Taking into account $\rho_{\mathrm{ary}}$ and considering (\ref{Eq:Sig_Gen}), the precoder matrix can be written as $\mathbf{F}=\frac{\lambda\sqrt{\rho_{\mathrm{ary}}}}{4\pi}\mathbf{DTB}$ which is given in (\ref{Eq:Precoder}) and concludes the proof.

\bibliographystyle{IEEEtran}
\bibliography{Ref_29_10_2018}

\end{document}